\documentclass[prl,preprint,amsmath,amssymb,nofootinbib,showkeys]{revtex4-1}

\usepackage{graphicx,amsmath,amssymb,times}
\usepackage{dcolumn}
\usepackage{bm}
\usepackage{amsmath}
\usepackage{array}
\usepackage{color}
\usepackage{amsmath}

\usepackage{scrextend}
\usepackage{longtable}

\usepackage{times}
\usepackage{tabularx}

\usepackage{array}

\begin{document}

\title{Precise nanoscale temperature mapping in operational microelectronic devices by {\color{black} use of a} phase change material}

\author{Qilong Cheng$^{1}$\footnote{Equal contribution.\label{refnote}}}
\author{Sukumar Rajauria$^{2}$\footref{refnote}}
\altaffiliation{email: sukumar.rajauria@wdc.com}
\author{Erhard Schreck$^{2}$}
\author{Robert Smith$^{2}$}
\author{Na Wang$^{2}$}
\author{Jim Reiner$^{2}$}
\author{Qing Dai$^{2}$}
\author{David Bogy$^{1}$}
\affiliation{$^{1}$Department of Mechanical Engineering, UC Berkeley, California 94720 USA.}
\affiliation{$^{2}$Western Digital Corporation, Recording Sub System Staging and Research, San Jose, CA 95135 USA.}

\begin{abstract}
\noindent
The microelectronics industry is pushing the fundamental limit on the physical size of individual {\color{black} elements} to produce faster and more powerful integrated chips. These chips have nanoscale features that dissipate power resulting in nanoscale hotspots {\color{black} leading to} device failures. To understand the reliability impact of  the hotspots, the device needs to be  tested under the actual operating conditions. {\color{black} Therefore, the development of high-resolution thermometry techniques is required to understand the  heat dissipation processes during the device operation.} Recently, several thermometry techniques have been proposed, such as radiation thermometry, thermocouple based contact thermometry, scanning thermal microscopy (SThM), scanning transmission electron microscopy (STEM) {\color{black}and transition based threshold thermometers}. However, most of these techniques have {\color{black}limitations} including {\color{black}the need for} extensive calibration, perturbation of the actual device temperature, low throughput, and {\color{black}the use of} ultra-high vacuum. Here, we present a facile technique, which uses a thin film contact thermometer based on the phase change material $Ge_2 Sb_2 Te_5$, {\color{black} to precisely map thermal contours from the nanoscale to the microscale}. $Ge_2 Sb_2 Te_5$ undergoes a crystalline transition {\color{black}{at T$_{g}$}} with {\color{black} large changes}  in its electric conductivity, optical reflectivity and density. Using this approach, we map the {\color{black} surface} temperature {\color{black} of a nanowire and an embedded micro-heater on the same chip} where {\color{black} the scales of the temperature contours} differ by three orders of magnitude. The spatial resolution can be as high as 20 nanometers thanks to the continuous nature of the thin film.
\end{abstract}


\date{\today}

\maketitle

 {\color{black} The}  fundamental understanding of thermal dissipation in an integrated chip  \cite{HaenschIBM06,CavinJNR06,PopNR10, DejeneNP13,JinNL11} requires the development of a versatile technique capable of reliably mapping the {\color{black} areal} temperature of various components integrated in the chip ranging from nanometer to micrometer dimensions  \cite{BritesNanoscale12, LeeSR07}. Various thermometers were developed to achieve this {\color{black}goal} \cite{WildeN06, DeshpandePRL09,TessierAPL07,BritesNanoscale12,NonnenmacherAPL92, ShiAPL92,MecklenburgScience15,BrintlingerNL08, GuoNC14, GaoN02,MartinekU15,MengesNC16,KinkhabwalaIEEE15,KimACS12,HwangRSI14} and {\color{black}can be} broadly classified into two categories: non-contact and contact {\color{black} based} thermometers. {\color{black} Techniques} such as Raman \cite{DeshpandePRL09}, fluorescence \cite{OkabeNC12}, luminescence \cite{KucskoNature13}  and transmission electron microscopy \cite{MecklenburgScience15, IdroboPRL18, FengNL18} are non-contact thermometers. However, the areal resolutions of these methods are {\color{black}limited either} by the optical diffraction limit \cite{ChristoffersonJEP08} or by the use of {\color{black}specific} metals and semiconductors \cite{MecklenburgScience15}. A scanning thermal microscope is an extensively used contact thermometer, {\color{black}but} {\color{black} it} typically suffers from contact-related artifacts {\color{black}that lead to} an under prediction of {\color{black}the} device temperature. This is due to {\color{black} the} thermal coupling strength between the device and the SThM tip, which is material dependent and difficult to measure \cite{MengesNC16, MengesNL12}.

%******************************************************************************************************
\begin{figure*}[htbp]
\begin{center}
\includegraphics[width=7in]{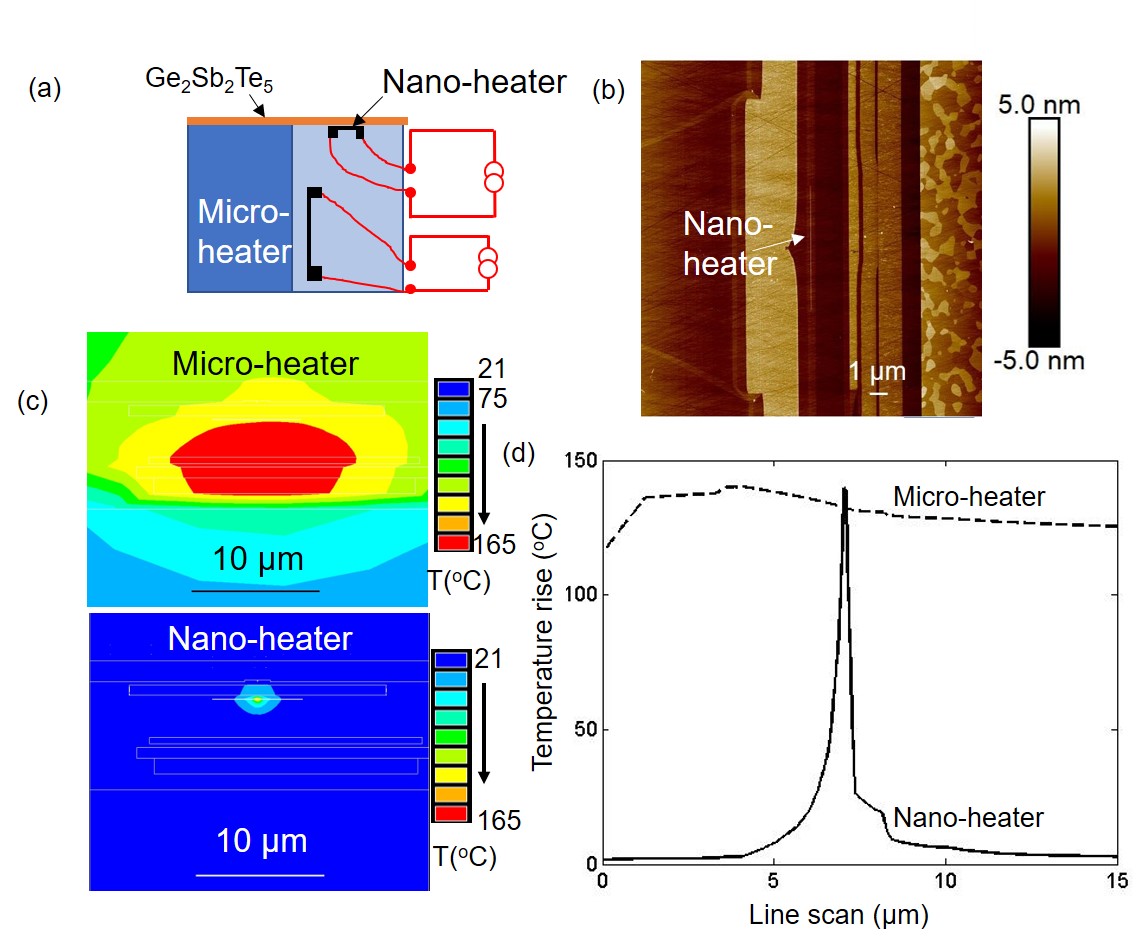}
\end{center}
\caption {$\textbf{Heat sources inside the {\color{black}head} of the hard disk drive}$. (a) Cross-sectional cartoon of the head structure showing the embedded heat sources: the nano-heater and the micro-heater. (b) AFM image of the device: the micro-heater is embedded and cannot be seen from the surface, while the nano-heater is {\color{black}located} at the center. The dimension of the nano-heater is 1 $\mu$m $\times$ 20 nm. (c), (d) Simulation: Temperature map of the nano-heater and the micro-heater with similar peak surface temperature. }
\label{fig:1}
\end{figure*}
%******************************************************************************************************

Here a novel technique, which uses a phase change material to map the temperature of an operational microelectronic device, is presented. It requires minimal effort in temperature calibration and {\color{black} the} temperature contour {\color{black} can}  be mapped using both {\color{black} contact}  and non-contact {\color{black}modes} such as AFM, SEM or optical microscopes \cite{FritzscheJNCS70, FriedrichJAP00, MoralesJAP02, LazarenkoAIP16, BurrJVST10}. We map the temperature contours of a nanowire and an embedded micro-heater where the contour areas differ by three order{\color{black}s}  of magnitude.

To demonstrate the versatility and practicality of this technique, a recording head from a commercial hard disk drive is used. The head of the hard disk drive provides a unique platform for such studies as it has several embedded heat sources, which differ in heated area by three orders of magnitude \cite{WuAPL16, ShimizuIEEE11}. At {\color{black} the}  microscale, it has {\color{black}a} micro-heater, which is used to adjust the clearance between the head and the rotating disk \cite{RajauriaPRL18}. {\color{black}The micro-heater} is embedded a few micrometers from the surface, and {\color{black} it produces}  a microscale temperature contour. At {\color{black} the}  nanoscale, it has a nano-heater, which is used both as a heater and a thermometer. {\color{black}The nano-heater} consists of a 200 nm {\color{black} wide}, 1 $\mu$m long, and 20 nm thick {\color{black}metal wire that is} embedded 2 nm from the surface. Figure 1(c) shows the simulated surface temperature contours for the micro-heater and {\color{black} the}  nano-heater. In comparison to the micro-heater{\color{black} 's temperature field}, the nano-heater{\color{black}'s} areal temperature map is three orders of magnitude smaller (see Figure 1(d)). {\color{black} Note} that the nano-heater {\color{black}has a} temperature coefficient of resistance (TCR) {\color{black}of 0.003/K, which} is used to measure the average surface temperature. {\color{black}In this paper, it is quantitatively demonstrated} that the temperature measured from the phase change temperature contour (PCTC) technique agrees well with the measured average surface temperature and the thermal simulation for both the micro-heater and the nano-heater.

\section*{Results and Discussions}

\subsection*{Self-heating of the nano-heater}

%******************************************************************************************************
\begin{figure*}[htbp]
\begin{center}
\includegraphics[width=6in]{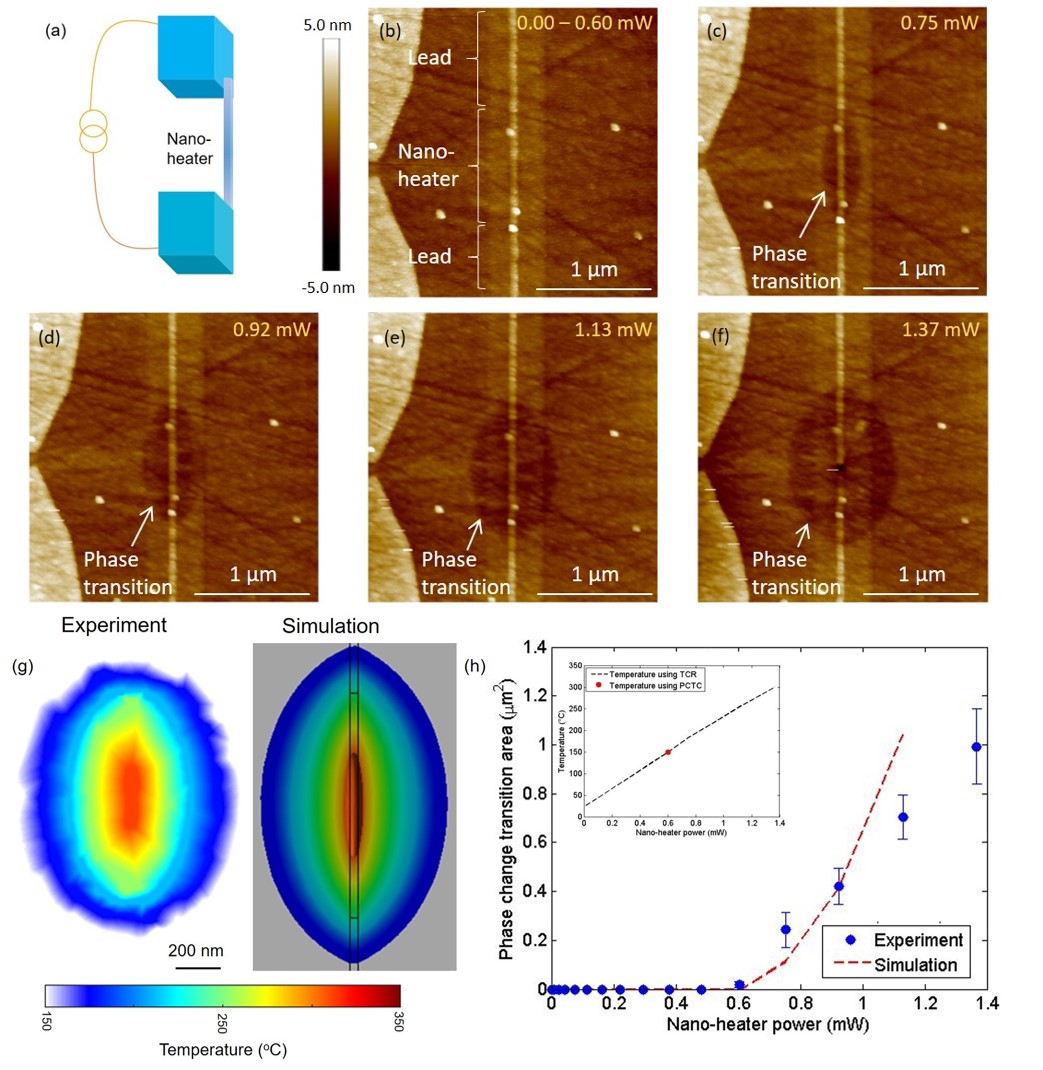}
\end{center}
\caption {$\textbf{Self-heating of the nano-heater}$.  (a) Schematic diagram of the nano-heater. Nanowire with dimension 1 $\mu$m $\times$ 20 nm $\times$ 200 nm is electrically connected to two pads. (b)-(f) {\color{black} Show} the AFM {\color{black} images} of the device at different micro-heater bias condition{\color{black} s}. (c)-(f) {\color{black} Show} the depression in the topography from the phase transition around the nano-heater. (g) Shows the constructed temperature contour from the PCTC technique and the simulation for the nano-heater power of 1.37 mW. (h) Shows the measured phase change transition area as a function of dissipation power in the nano-heater power. The red dash line corresponds to the simulation of an isotherm contour for the glass transition temperature T$_g$.  In the inset, the black dash line shows the estimated average surface temperature along the nano-heater from the resistance change in the nano-heater and the measured isotherm from the PCTC technique (Red dot). Estimated error bar in average surface measurement is 0.04 $^{o}$C. }
\label{fig:2}
\end{figure*}
%******************************************************************************************************

%******************************************************************************************************
\begin{figure*}[htbp]
\begin{center}
\includegraphics[width=7in]{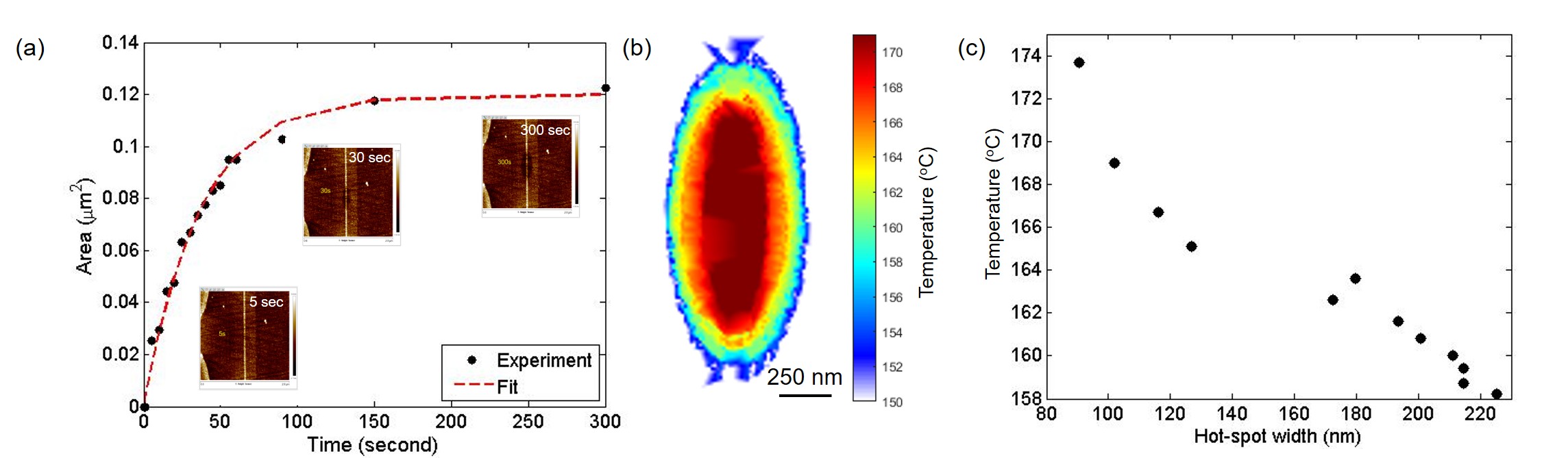}
\end{center}
\caption {{\color{black}$\textbf{Time response of $Ge_2 Sb_2 Te_5$ at a constant nano-heater {\color{black} power of 0.68 mW}.}$  (a) The transition area of $Ge_2 Sb_2 Te_5$ phase change with the {\color{black} the accumulated heating time}. (b) Shows the temperature map of the nano-heater. (c) Shows the temperature as a function of {\color{black}the} hotspot width {\color{black} across} the nano-heater.}}
\label{fig:2_1}
\end{figure*}
%******************************************************************************************************

To demonstrate the technique, we first characterize the Joule heating of {\color{black}the} nano-heater inside the {\color{black}head} of the hard disk drives. The {\color{black}head surface} is coated with a {\color{black} 22 nm thick layer of} {\color{black}$Ge_2 Sb_2 Te_5$  thin film.} {\color{black} The} nano-heater is biased using a current source, {\color{black} across which the measured voltage drop} is used to estimate {\color{black} the resistance increase of the nano-heater due to the dissipated Joule heat} (Figure 2(a)). {\color{black} The} resistance change of the nano-heater is used to estimate the average temperature increase {\color{black} of the} device temperature {\color{black}by} $R_{T}=R_{0}(1+\alpha \Delta T)$, where {\color{black} $R_{0}$ is the room-temperature resistance at low current bias where no significant self-heating occurs, $R_{T}$ is the resistance at the bias corresponding to the temperature T,} $\alpha$ is the temperature coefficient of resistance (TCR), and $\Delta$T is the average temperature rise due to {\color{black}the} Joule {\color{black} heating}. The temperature coefficient of resistance ($\alpha$) 0.003/K is determined separately in an oven using a 4-probe measurement scheme (see Supplementary {\color{black}section 1}). {\color{black} Note that the effect of the thin layer on the heat transport of the system is negligible (see Supplementary {\color{black}section 2}).}

The amorphous $Ge_2 Sb_2 Te_5$ is a chalcogenide phase change material that crystallizes {\color{black}at} T$_{g}\sim149^{\circ}$C {\color{black} for a dwell time of 5 minutes}. This crystallization is accompanied by an increase in density and volume reduction, {\color{black} where AFM topography measurement shows} as a reduction in {\color{black} the} film height. Figures {\color{black}2(b-f)} {\color{black} show}  the AFM topography micrograph{\color{black} s} {\color{black} corresponding to} different {\color{black} powers} in the nano-heater. For nano-heater power smaller than {\color{black}0.60} mW, the AFM shows no change in the topography of the $Ge_2 Sb_2 Te_5$ film over the nano-heater, indicating that the surface temperature is lower than the crystallization temperature. {\color{black} When the nano-heater power is 0.75} mW, a small depression in the topography {\color{black}is observed centered at the hot spot of the nano-heater as shown in Figure 2(c).} {\color{black} A} further increase in the nano-heater power leads to a gradual increase in the area undergoing the crystallization{\color{black}, which} is indicated by the lateral growth of the depressed area in the AFM {\color{black}images}. {\color{black} Note} that the boundary of the topography depression corresponds to the isotherm {\color{black}of} the crystallization temperature. The evolution of the temperature contour area agrees reasonably well with the simulation {\color{black} as shown in Figure 2(h). Furthermore,} {\color{black}we use the transition boundary measured at different nano-heater powers to map the temperature of the device. {\color{black} The} rate of phase transition in $Ge_2 Sb_2 Te_5$ is a function of both the temperature and the time. Here, the power in {\color{black} the} nano-heater is increased incrementally with a fixed dwell time until the initial transition boundary is observed. {\color{black}The last transition boundary corresponds to the calibration temperature T$_{g}$ at the largest heater power P$_{o}$=1.37 mW (Figure 2(f)). Assuming that the temperature is linear with the applied power, the temperature isotherm T$_{i}$ at each previous transition boundary (Figures 2(c-e)) is given by:}
\begin{equation}
T_{i}=T_{g}\frac{P_{o}}{P_{i}}
\label{eq:1}
\end{equation}
where, T$_{g}$ is the calibrated $Ge_2 Sb_2 Te_5$  crystalline transition temperature for the dwell time of 300 seconds {\color{black} during} which {\color{black} the} nano-heater is powered on,  P$_{o}$ is the nano-heater power at which the {\color{black}final} transition boundary is measured, and  P$_{i}$ is the {\color{black}previous} power with  P$_{i}$ {\color{black}$<$} P$_{o}$ in the nano-heater. Figure 2(g) shows the constructed temperature map of the device along with {\color{black} the} simulation {\color{black}for P$_{o}$=1.37 mW.}}

%******************************************************************************************************
\begin{figure*}[htbp]
\begin{center}
\includegraphics[width=7in]{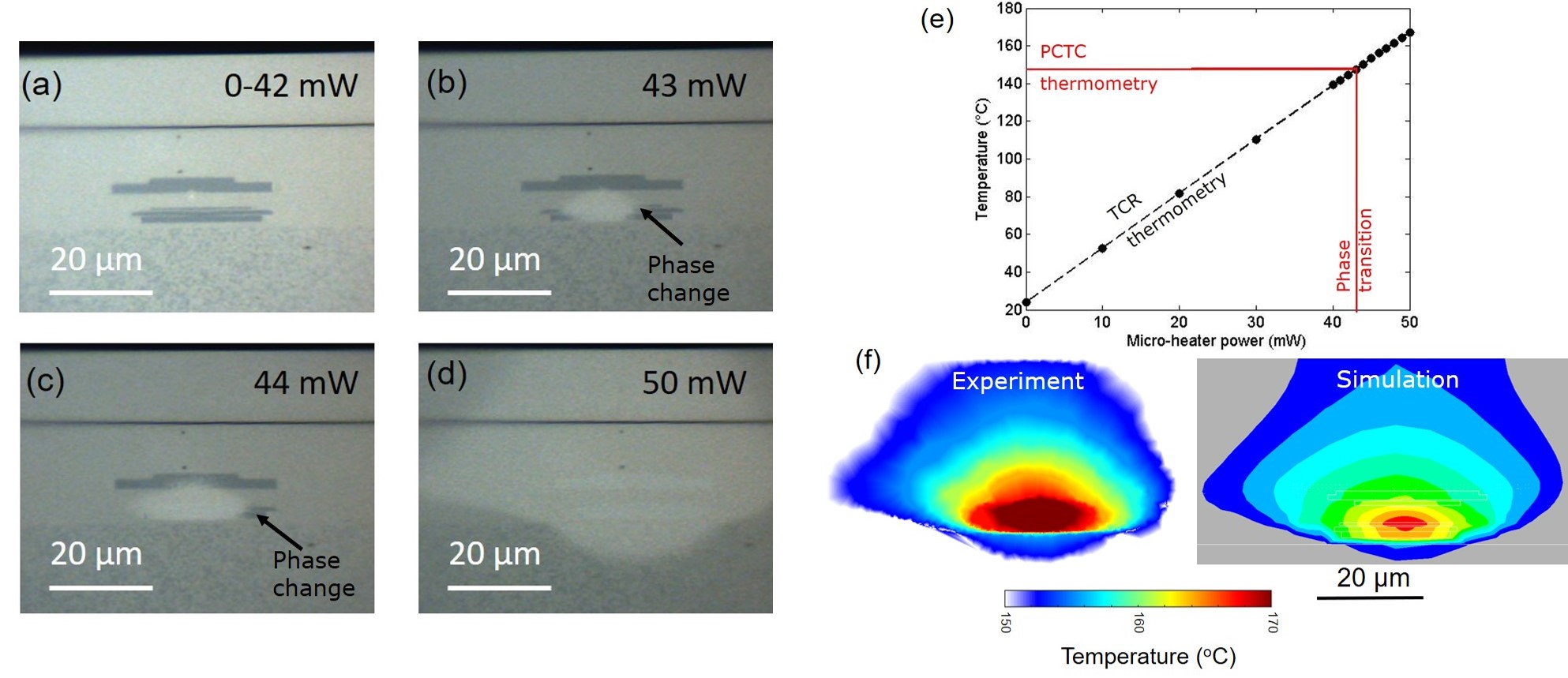}
\end{center}
\caption {$\textbf{Self-heating of the micro-heater}$. (a)-(d) {\color{black}Show} optical micrograph{\color{black} s} of the device at different micro-heater powers. (b)-(d) {\color{black}Show} the reflectivity increase at the center of the {\color{black}micrographs} corresponding to the phase change {\color{black}due to the temperature rise of the micro-heater.} (e) Shows the average surface temperature measured by the nano-heater, acting as a thermometer, along with the {\color{black} critical point} for which {\color{black}the phase transition} is measured from the optical micrograph. \color{black}{Estimated error bar in average surface measurement is 0.04 $^{o}$C.} (f) Shows the {\color{black}constructed} temperature map of the device along with the simulation at the micro-heater power of 50 mW.}
\label{fig:3}
\end{figure*}
%******************************************************************************************************

Figure 2(h) shows the phase change transition area calculated from the topography depression in the AFM image as a function of the power dissipated in the nano-heater. {\color{black}When the nano-heater power is lower than 0.60 mW}, the transition area is zero signifying that the surface temperature is lower than the glass transition temperature {\color{black}T$_{g}$ everywhere. At} higher powers the transition area grows linearly with the dissipated power in the nano-heater. {\color{black} The} simulation results are shown as the red dash line. Both the experiment and the simulation show a sharp increase in the phase change transition area beyond 0.60 mW. At much larger bias currents the poor match is {\color{black}because of} our inability to capture the exact structural details in the simulation {\color{black} such as the actual thermal boundary conditions and the various material parameters}. {\color{black} To further} confirm the surface temperature, we simultaneously {\color{black}measure} the resistance of the nano-heater and {\color{black} use a }  TCR of 0.003/K to estimate the average temperature of the heater. Figure 2(h) inset shows the measured surface temperature of the nano-heater as a function of the power {\color{black} dissipated}. {\color{black} The} red dot is the temperature from the PCTC technique at 0.60 mW {\color{black} {\color{black} derived from the} x-axis intercept of the Figure 2(h). As expected, the temperature from the PCTC technique matches the temperature reading given by the resistance change}. The excellent agreement between the temperature measured using the PCTC and TCR technique along with the simulation demonstrates that {\color{black} the} PCTC technique can precisely {\color{black}map} the high operational temperature of the nanoscale {\color{black}heater} embedded in the chip.

{\color{black} {\color{black}{\color{black} Next} we study} the time response of the $Ge_2 Sb_2 Te_5$ film to construct the temperature map of the device at a constant nano-heater power. Here, the growth of the transition boundary is tracked over time. It should be noted that {\color{black}the transient response of the nano-heater} is six orders of magnitude faster than {\color{black} the} {\color{black}$Ge_2 Sb_2 Te_5$} {\color{black} phase change}. Assuming that the phase change conversion follows an Arrhenius model {\color{black}and the conversion is linear with time}, the temperature for each transition boundary is derived by monitoring the time needed for each transition boundary to develop. The temperature ($T_{i}$) at time $t_{i}$ is given by:
\begin{equation}
T_{i}(t_{i})=\left[-\frac{k_{B}}{E_{A}}\left( ln\frac{1}{t_{i}} -  ln\frac{1}{t_{cal}} \right) + \frac{1}{T_{g}(t_{cal})}  \right]^{-1}
\label{eq:2}
\end{equation}
where, $k_{B}$ is {\color{black} the} {\color{black} Boltzmann} constant, $E_{A}$ $\sim$ 2.6 eV is the activation energy $Ge_2 Sb_2 Te_5$ transition, T$_{g} \sim$ 149$^{\circ}$C is the calibrated {\color{black}crystallization} temperature {\color{black} at dwell time} t$_{cal}$ =300 sec. It is worth noting that the temperatures of isotherms corresponding to shorter dwell {\color{black}times} ($t_{i}$ $<$ $t_{cal}$) are higher than T$_{g}$.

Figure 3(a) dots {\color{black}show} the phase change transition area around the nano-heater as a function of {\color{black}the accumulated} time for a constant power of {\color{black}0.68} mW (see Supplementary {\color{black}section 3} and video TimeDependence.mp4). {\color{black} The} red line is the exponential fit with a time constant of 37.6 seconds. {\color{black} The transition temperature at different accumulated heating time is determined by using Eq. 2.} Figure 3(b) shows the constructed temperature map of the device at {\color{black}the constant nano-heater power of 0.68 mW}. In comparison to Figure 2(g), the temperature map is smaller and {\color{black} more} elliptical {\color{black} since the} nano-heater power is almost 50$\%$ {\color{black} smaller}. Figure 3(c) shows the temperature across nano-heater as a function of distance demonstrating the high resolution of the PCTC scheme. {\color{black} The} continuous nature of our {\color{black}thin film} allows for a higher spatial resolution, which {\color{black} is limited only by the grain size of $Ge_2 Sb_2 Te_5$ (sub 20 nm) and the resolution of the imaging technique ($<$ 10 nm) \cite{BrintlingerNL08}.}}

\subsection*{Self-heating of the micro-heater}

To demonstrate the versatility of the PCTC technique, we now characterize the Joule heating of a much larger micro-heater embedded in the {\color{black}head} of the hard disk drive. {\color{black} The temperature contour of the micro-heater} is three orders of magnitude larger than {\color{black} that of} the nano-heater embedded in the same chip ({\color{black}as} shown in Figure 1) {\color{black} in terms of the {\color{black} contour} area}. {\color{black} The} micro-heater is biased using a current source and the measured voltage drop across the {\color{black}nano-heater (thermometer)} is used to estimate the dissipated Joule heat. The dwell time of 300 {\color{black}seconds} at a constant micro-heater power is much longer than the thermal response time of the heater and the phase transition time beyond which the physical, optical and electrical properties {\color{black}change}. {\color{black} The temperature contour of the micro-heater} is mapped using an optical microscope by simply imaging the reflectivity change in the transition area. Figures 4(a-d) show the optical micrographs at different micro-heater power{\color{black} s}. No change in reflectivity is observed {\color{black} below} {\color{black}the dissipated power of 42 mW} in the micro-heater. At the power of 43 mW, {\color{black} an} increase in the reflectivity is observed at the center of {\color{black} the} thermal hotspot {\color{black}due to} the micro-heater. {\color{black} Note} that the {\color{black}boundary} of the transition area corresponds to the crystallization temperature. In comparison to the nano-heater, here the micro-heater requires 60 times more power to achieve the same surface temperature {\color{black}since the micro-heater is more deeply embedded and heats up a much larger volume}. At higher micro-heater power{\color{black} s}, the growth in the transition area indicate{\color{black} s} an increase in {\color{black} the} thermal spot size with the same crystallization temperature T$_{g}$ isotherm.

Figure 4(f) shows the constructed temperature map of the device along with the simulation. {\color{black} T}he {\color{black}dimensions} and the overall shape of the transition contour from the {\color{black}experiments} match well with the simulated T$_{g}$ {\color{black}isotherms}. {\color{black} Furthermore, we utilized the nano-heater as a thermometer by monitoring its resistance change at a very low current of 0.1 mA, in order to avoid self-heating, to measure the temperature rise due to the micro-heater.} Figure 4(e) shows the surface temperature measured by the nano-heater as a function of the power dissipated in the micro-heater. The red line shows the micro-heater power beyond which the phase transition is observed in the optical micrograph. The expected {\color{black}rise of the surface temperature} as derived from both the PCTC technique and from the nano-heater (acting as 'thermometer') is 2.9 K/mW. This shows an excellent agreement between the PCTC technique, the measured surface temperature and the simulation for the temperature map of the micro-heater.

%******************************************************************************************************
\begin{figure}[htbp]
\begin{center}
\includegraphics[width=3.5in]{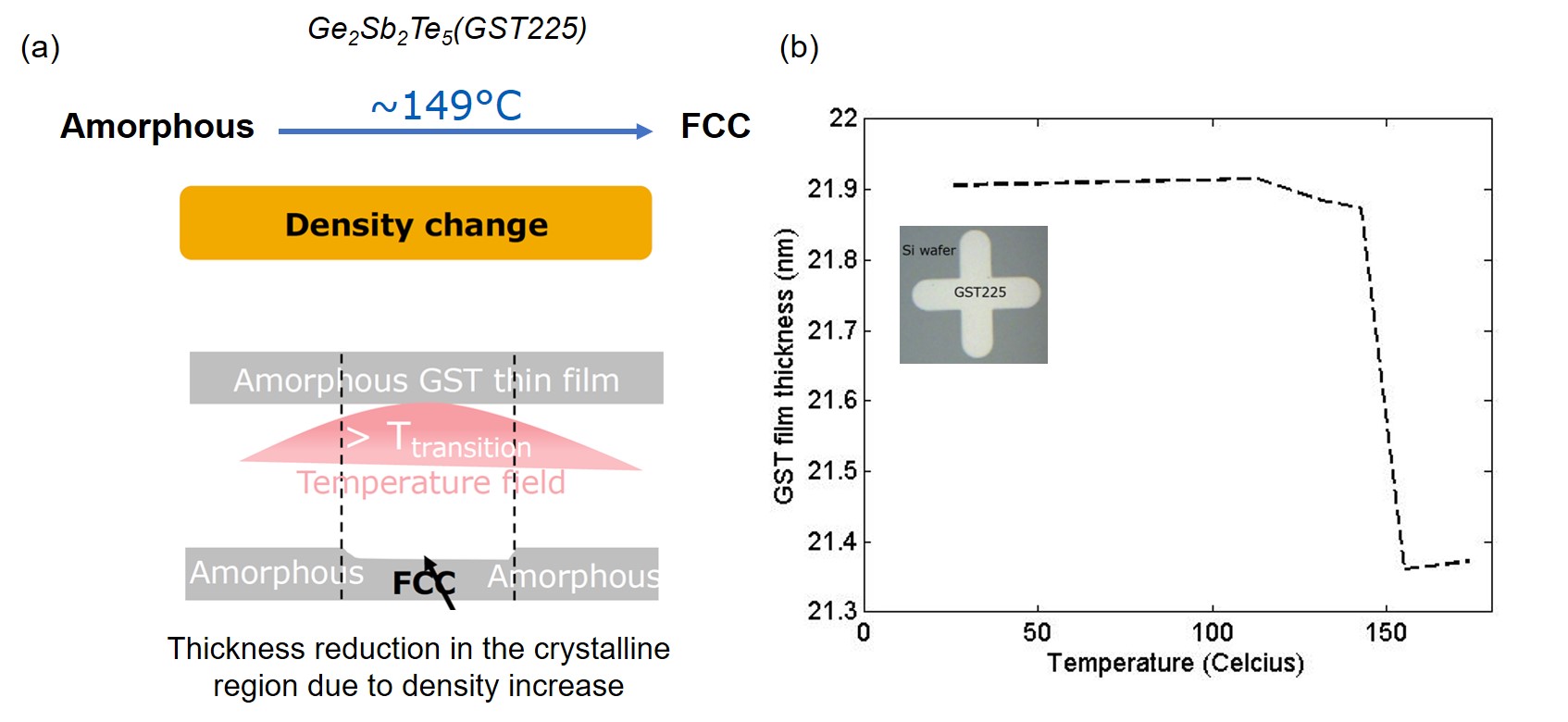}
\end{center}
\caption {$\textbf{Calibration}$. (a) Cartoon to show the phase transition in $Ge_2 Sb_2 Te_5$. (b) Inset: Optical micrograph of the calibration sample. Main: Thickness (measured using AFM, with vertical resolution of 0.05 nm) of the calibration sample as a function of {\color{black}oven} temperature. The glass transition temperature T$_{g}$ is 149$^{\circ}$C.}
\label{fig:4}
\end{figure}
%******************************************************************************************************

\subsection*{Temperature calibration}

Precise and relatively simple temperature calibration is a key advantage of the PCTC method compared to other techniques that require extensive temperature calibration. To calibrate the crystallization temperature T$_{g}$ of {\color{black} the} phase change material $Ge_2 Sb_2 Te_5$, we rely on {\color{black} a} pronounced structural property change at the phase change condition (Figure 5(a)). The rate of this transformation from an amorphous {\color{black}state} to the crystalline rock salt structure is well characterized by an activation energy of about 2.6 eV \cite{FriedrichJAP00, MoralesJAP02}. Due to this activation energy driven process the crystallization temperature {\color{black}is dependent} on the dwell time of the sample. For this reason, the dwell time at a constant power condition {\color{black}is} fixed at 5 minutes for both the nano-heater and the micro-heater. This dwell time is much longer than the response time of the two heaters \cite{SchreckIEEE92}. Figure 5(b) inset shows the calibration sample, which is {\color{black} a} photo-lithography defined 22 nm thick $Ge_2 Sb_2 Te_5$ layer on top of a silicon wafer. {\color{black} The} main figure shows the film thickness as a function of oven temperature. The dwell time at a constant oven temperature {\color{black}is} 5 minutes after which the sample {\color{black}is} allowed to cool down to room temperature, and the film thickness {\color{black}is} measured using the AFM. At T = 149 $^{\circ}$C, the film thickness reduces, indicating a phase transition from the amorphous state to the crystalline state, as confirmed by the X-ray diffraction pattern (see Supplementary {\color{black}section 4}).

The uncertainty in the temperature {\color{black} derived} from this {\color{black}PCTC} technique is primarily due to the fact that the crystallization rate of the phase change material does not have a large abrupt jump at a single temperature (as in a first order phase transition). As a result, the full temperature history of the sample, not just the last power used, can influence the size of the observed contour (see Supplementary {\color{black}section 5}). For both the nano-heater and the micro-heater, the estimated temperature step {\color{black}is} kept at 10 K, which leads to a slight under-prediction of {\color{black} the} temperature by around {\color{black}2 K} at the 149 $^{\circ}$C crystallization condition.

It is noteworthy that in most other {\color{black} high resolution} {\color{black} temperature mapping} techniques, {\color{black} the {\color{black} application} is limited} {\color{black} to the large scale} {\color{black}devices}. Here {\color{black}we demonstrated the thermal measurements of the two heaters} that produced thermal contours with dimensions that differ by three orders of magnitude, while maintaining sub {\color{black}20} nm resolution in both cases. This technique is extremely versatile and {\color{black} does not require} the use of expensive microscopes like {\color{black}STEM}. At {\color{black}the} microscale, even an inexpensive optical microscope can be used to map the temperature {\color{black}of} the hotspots or heat sources in operating microelectronics devices.

Finally, {\color{black}the limitations of the presented PCTC technique are discussed}. Although the technique is versatile and {\color{black} can} be used for nanoscale to microscale spatial heat sources with minimal calibration challenges, it has two main limitations. First, the {\color{black} areal} contour represents the isotherm at T$_{g}$, and to extract the temperature gradient one needs to perform mapping for at least two different power levels {\color{black} or track the phase change with time}. Secondly, the technique is limited to {\color{black} a} device temperature higher than the crystallization temperature of the deposited phase change material. In our case, {\color{black}$Ge_2 Sb_2 Te_5$ crystallizes at temperature of 149 $^{\circ}$C {\color{black} for the dwell time of 5 minutes} is used}. Both embedded heaters {\color{black}are} able to reach temperatures higher than this T$_{g}$. For other systems where reaching a similar temperature would be difficult or impossible, {\color{black}the issue} could be overcome by choosing {\color{black}a different composition or} phase change materials with lower T$_{g}$ \cite{XiongIEEE16,ChenIEEE98,FuRSC15,AlkanSE12}.

\section*{Conclusion}

To summarize, we {\color{black}introduce} a versatile {\color{black}phase-change-material-based} temperature mapping technique for operational microelectronic devices {\color{black} that can} spatially resolve temperature from nanoscale to microscale dimensions. It can be used to characterize surface {\color{black}temperatures with neglectable temperature interference due to the deposited measurement film and  with minimal} calibration. A thorough understanding of the heat dissipation in various nanoscale devices, such as {\color{black}the aforementioned nano-heater}, may lead to more efficient and powerful integrated chips, and hence holds great economic value to the industry.

\section*{Methods}
\subsection*{Experimental set-up}
The microelectronic device is held on a metal fixture with electrical pins. The components inside the device such as the heaters are powered by Keithley 2602 SYSTEM SourceMeter, which is controlled by a Python script. A 22 nm thin film of Ge2Sb2Te5 is sputtered on the surface of the device. The topography change or the reflectivity change of the thin film due to the heaters are measured by Digital Instruments Dimension 3100 AFM or an optical microscope respectively. In the temperature calibration of the Ge2Sb2Te5, a silicon wafer with a photo-lithography defined 22 nm thick Ge2Sb2Te5 layer is heated in a customized copper chamber, where the temperature is measured by a type-K thermocouple. The thickness of the layer is also measured by the AFM. 

\subsection*{Simulation}
Thermal simulations of the micro-heater and nano-heater devices in Figure 1 (c) and 1(d) were performed using finite element models in ANSYS Mechanical APDL version 17.2. The nano-heater thermal simulation in Figure 2 (g) was modeled using a finite element model in the ANSYS Workbench Thermo-electric module version 17.2. The device surfaces were modeled using a convection cooling boundary condition with a coefficient of 50 W/(m$^{2}$·K).    

%****************************ACKNOWLEDGEMENTS**************************************************************
\section{Acknowledgements}
%\textbf{Acknowledgements}
%\\
The authors would like to thank I. McFadyen for careful reading of the manuscript. Q.C would like to thank Western Digital Corporation for financial support for the internship.

%**********************************CONTRIBUTIONS**************************************************************
\section{Author Contributions}
The work was conceived by S.R. and E.S. The experimental setup were designed and implemented by Q.C. and S.R. The experiments were performed and analyzed by Q.C. under the supervision of S.R. and E.S. Thermal simulation was done by R.S. Phase change material was deposited by J.R. The manuscript was written by S.R. and Q.C with comments and
input from all authors.
\section{Author information}
Corresponding author: sukumar.rajauria@wdc.com

The authors declare they have no competing financial interests.

 \end{document}